\documentclass[aps,12pt,showpacs,nofootinbib,superscriptaddress,epsfig]{revtex4}
\usepackage{graphicx}
\usepackage{dcolumn}
\newcommand{\be}{\begin{equation}}
\newcommand{\ee}{\end{equation}}
\newcommand{\bea}{\begin{eqnarray}}
\newcommand{\eea}{\end{eqnarray}}
\newcommand{\beas}{\begin{eqnarray*}}
\newcommand{\eeas}{\end{eqnarray*}}
\newcommand{\bd}{\begin{displaymath}}
\newcommand{\ed}{\end{displaymath}}
\newcommand{\ii}{\'{\i}}
\def\shiftleft#1{#1\llap{#1\hskip 0.04em}}
\def\shiftdown#1{#1\llap{\lower.04ex\hbox{#1}}}
\def\thick#1{\shiftdown{\shiftleft{#1}}}
\def\b#1{\thick{\hbox{$#1$}}}

\begin{document}

\title{ Axial $N\to \Delta(1232)$ and $N \to N^{\star}(1440)$ transition form factors}

\bigskip

\author{D. Barquilla-Cano}\thanks{Email:dbcano@usal.es}\affiliation{Grupo de F\'{\i}sica Nuclear, 
Departamento de F\ii sica Fundamental e 
IUFFyM, Facultad de Ciencias, E-37008 Salamanca, Spain}   
\author{A. J. Buchmann}\thanks{Email:alfons.buchmann@uni-tuebingen.de}
\affiliation{Institut f\"ur 
Theoretische Physik, Universit\"at T\"ubingen, 
Auf der Morgenstelle 14, 72076 T\"ubingen, Germany} 
\author{ E. Hern\'andez}\thanks{Email:gajatee@usal.es}
\affiliation{Grupo de F\'{\i}sica Nuclear, 
Departamento de F\ii sica Fundamental e 
IUFFyM, Facultad de Ciencias, E-37008 Salamanca, Spain}

\vskip2cm
\begin{abstract}
We calculate the axial $N\to \Delta(1232)$ and $N\to N^{\star}(1440)$ 
transition form factors in a chiral constituent quark model. 
As required by the partial conservation of axial current ($PCAC$) condition, we 
include one- and two-body axial exchange currents. 
For the axial $N\to \Delta(1232)$ form factors we compare with previous quark
model calculations that use only one-body axial currents, 
and with experimental analyses. The paper provides the first calculation 
of all weak axial $N\to N^{\star}(1440)$ 
form factors. Our main result is that exchange currents are very important
for certain axial transition form factors. 
In addition to improving our understanding of nucleon structure, 
the present results are relevant for neutrino-nucleus scattering cross section 
predictions needed in the analysis of neutrino mixing experiments.

\end{abstract}

\maketitle

\section{Introduction}
\nobreak

The axial $N \to N^{\star}$ transition form factors play an important role 
in neutrino induced pion production on the nucleon, e.g. in 
${\bar \nu_e} + p \to \Delta^0 + e^+ \to n + \pi^0 + e^+$.            
The two lowest-lying nucleon resonances, $\Delta(1232)$ and 
$N^{\star}(1440)$ (Roper resonance) are expected to give the dominant contribution to 
the neutrino scattering 
cross section for moderate neutrino energies. 
Weak $\Delta(1232)$ production has been studied 
experimentally in a series of neutrino scattering experiments on hydrogen and deuterium
targets~\cite{radecky,barish,kitagaki}. New data on the $N \to \Delta$ axial vector 
transition form factor are expected from experiments at Jefferson Laboratory~\cite{latifa}.
On the theoretical side, the  weak axial $\Delta$ excitation 
has been attracting attention since the 1960's and has been studied using different
approaches. For an overview see Refs.~\cite{schreiner,c5q3}. 
The first lattice computation of $N\to\Delta$ axial form factors has just
appeared~\cite{lattice}. 


To our knowledge there is no experimental information on the axial 
$N \to N^{\star}(1440)$ transition form factors. In Ref.~\cite{luis98}
the authors provided a theoretical estimate of the weak $N^{\star}(1440)$ production cross section 
in electron induced reactions 
in the kinematic region of the $\Delta$ resonance but no prediction for
the axial $N \to N^{\star}(1440)$ form factors was made.
The only theoretical determination of weak form factors for the 
$N \to N^{\star}(1440)$ transition that we are aware of,
was done in Ref.~\cite{bruno} but there only one of them, namely $g_A^\star$ (see below) 
was evaluated.

It is important to have quark model predictions for the weak $N  \to N^{\star}$ transition 
form factors for two reasons. First, they contain 
information on the spatial and spin structure of the nucleon and its excited states that 
is complementary to that obtained from electromagnetic $N\to N^{\star}$ form factors~\cite{mec}.
Second, they are required for neutrino-nucleus scattering cross section predictions 
which in turn are needed for a precise determination of neutrino mass differences 
and mixing angles~\cite{Pas04}.
Previous quark model calculations~\cite{c5q3,Abd72,c5q1,c5q1-2,c5q2,c5q4,c5q5}
included only one-body axial currents, i.e. processes where 
the weak probe couples to just one valence quark at a time (impulse approximation). 
However, this approximation violates the partial conservation of axial current (PCAC) condition, 
which requires that the axial current operator be a sum of one-body and two-body exchange terms,
and that the latter be connected with the two-body potentials of the quark model 
Hamiltonian~\cite{david1,david2}. The axial exchange currents provide an effective description 
of the non-valence quark degrees of freedom in the nucleon as probed by the weak interaction. 

\begin{figure}
{\resizebox{15cm}{5cm}{\includegraphics{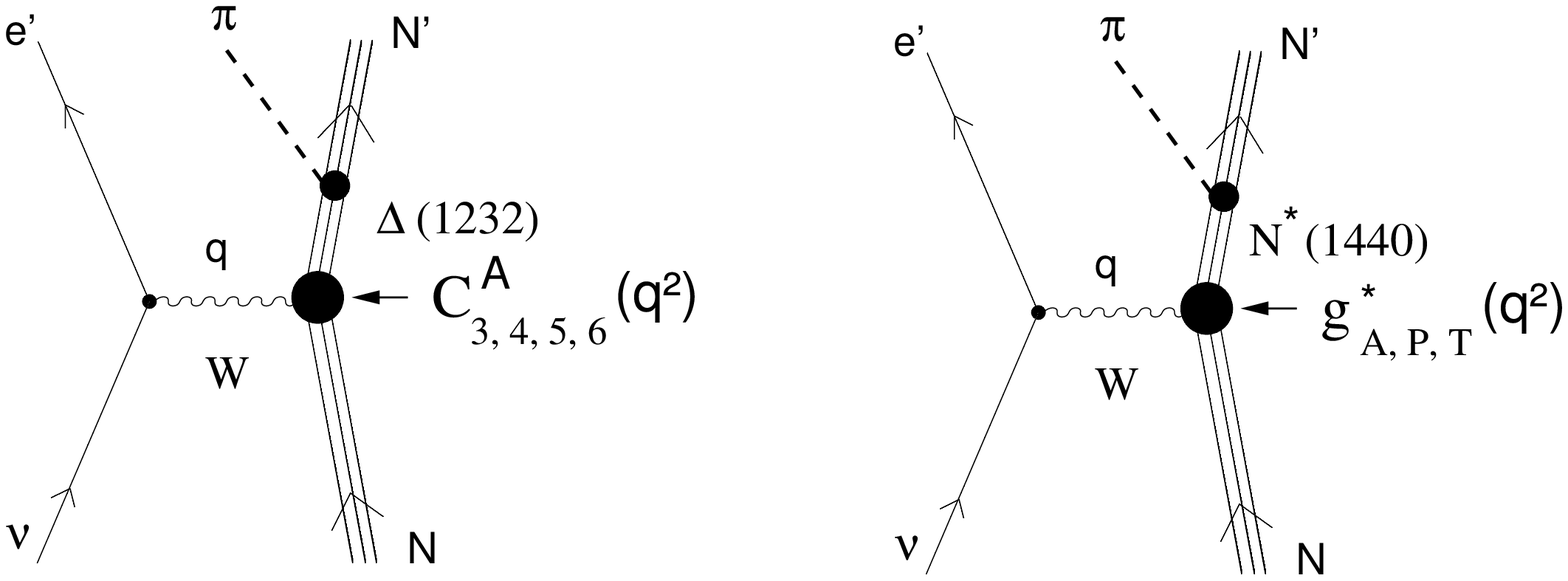}}}
\caption{Feynman diagrams for neutrino-induced pionproduction on the nucleon via
resonance excitation.
 Left: Axial transition form factors 
$C_{3}^A(q^2)$, $C_{4}^A(q^2)$, $C_{5}^A(q^2)$, and $C_{6}^A(q^2)$ 
contributing 
to weak $N \to \Delta(1232)$ excitation.
Right: Axial transition form factors 
$g_{A}^*(q^2)$,  $g_{P}^*(q^2)$, and $g_{T}^*(q^2)$ contributing to 
weak excitation of the $N^*(1440)$ resonance. The four-momentum transfer of 
the weak gauge boson $W$ is denoted by $q$.  }
\label{fig:feynman} 
\end{figure}

Recently, employing a chiral quark model with gluon and pseudoscalar meson exchange potentials and 
corresponding axial exchange currents, we have evaluated the elastic axial nucleon 
form factors $g_A(q^2)$ and $g_P(q^2)$~\cite{david1}, 
as well as the axial couplings $g^8_A(0)$ and $g^0_A(0)$
related to the spin content of the nucleon~\cite{david2}. 
The results obtained were in good agreement with experiment. Furthermore,
they allowed a consistent quark model interpretation of the missing nucleon spin
as orbital angular momentum carried by the nonvalence quark degrees of freedom
in the nucleon.  In the present paper we apply this model to the weak excitation of nucleon 
resonances as shown in Fig.~\ref{fig:feynman} and
calculate the axial $N\to \Delta(1232)$ and $N \to N^{\star}(1440)$ 
form factors.
As in our previous work, we go beyond the impulse approximation and 
include not only pion exchange currents but also two-body axial currents arising from gluon 
exchange and the confinement interaction as required by PCAC.
We will see below that in certain axial form factors, the contribution of various 
exchange currents can be clearly identified, and thus further details of nucleon structure 
can be revealed. 
 
The paper is organized as follows. 
After a short review of the chiral quark model in sect.~2, 
we calculate in sect.~3 all four Adler
form factors $C^A_i(q^2), \ i=3 \cdots 6$ of the weak $N \to \Delta(1232)$ transition,
and compare with other theoretical calculations and experimental analyses of
neutrino-induced pionproduction on the nucleon.
Sect.~4 is devoted to the axial $N \to N^{\star}(1440)$ transition, for which we 
present the first theoretical prediction of all three axial form factors. 
We summarize our results in sect.~5.

\section{Chiral quark model} 

The calculation of the axial form factors is performed within the framework of
the chiral constituent quark model in which chiral symmetry is introduced via 
the non-linear $\sigma$-model.
Although we refer the reader to Ref.~\cite{david2} for details of the model,
we explain here its main ingredients. The Hamiltonian includes apart from a 
confinement potential $V^{conf}$, a one-gluon exchange potential $V^{g}$, 
and a one-pion exchange potential $V^{\pi}$
~\footnote{ For the observables calculated here, 
the contribution of the $\eta_8$ exchange potential and axial current 
is small and can be ignored.}
\be
\label{hamiltonian}
H=\sum_{j=1}^{3}\left( m_q+\frac{\b{p}_j^2}{2m_q}\right)-
\frac{\b{P}^2}{6m_q}
+\sum_{j<k=1}^{3}\left(
V^{conf}({\bf r}_j, {\bf r}_k)  
+V^{g}({\bf r}_j, {\bf r}_k)  
+V^{\pi}({\bf r}_j, {\bf r}_k)  \right ), 
\ee
where $m_q$ is the constituent quark mass. Here, 
${\bf r}_j$, ${\bf p}_j$ are the position and momentum operators 
of the $j$-th quark, and ${\bf P}$ is the momentum of 
the center of mass of the three-quark system.
The kinetic energy of the center of mass motion is subtracted from
the total Hamiltonian. Explicit expressions for the individual potentials 
can be found in Ref.~\cite{david2}.

\begin{figure}
{\resizebox{14cm}{4cm}{\includegraphics{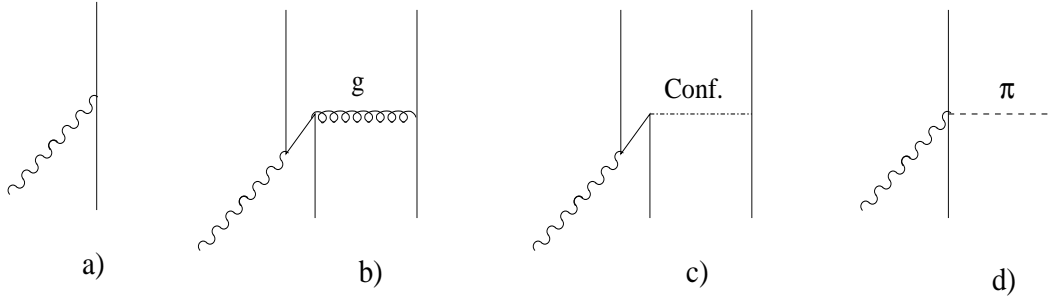}}}
\caption{Feynman diagrams for the axial currents employed in this work: 
a) one-body current (impulse approximation), b)
one-gluon exchange current, c) confinement exchange current,  d) one-$\pi$ exchange current.}
\label{fig:model} 
\end{figure}

The axial currents employed in this work are shown in Fig.~\ref{fig:model}. 
As mentioned in the introduction, the axial current operator contains not only one-body currents 
but also two-body gluon, pion, and confinement exchange currents consistent
with the two-body potentials in Eq.(\ref{hamiltonian}) as required by the PCAC relation
\be
\label{pcac}
{\bf q} \cdot {\bf A}({\bf q}) -
[H,A^0({\bf q})]
= -i\ \sqrt2\ f_{\pi}\frac{m_{\pi}^2}{q^2-m_{\pi}^2} M^{\pi}({\bf q}).
\ee
The PCAC equation links the strong interaction Hamiltonian
$H$, the weak axial current $A^{\mu}=(A^0,{\bf A})$ operators,
and the pion emission operator described by $M^{\pi}$. Here,
$m_{\pi}$ is the pion mass and $f_{\pi}$
is the pion decay constant.
Eq.(\ref{pcac}) also demands 
that the axial coupling of the quarks, $g_{Aq}$, is related to the pion-quark coupling constant, 
$g_{\pi q}$, via a Goldberger-Treiman relation~\cite{david1}
\be
\label{gtrquark}
g_{Aq}=f_{\pi}\frac{g_{\pi q}}{m_q}.
\ee
Inserting the physical values for the constituent quark mass, the pion decay constant,
and the pion-quark coupling, 
one finds  that $g_{Aq}$ is renormalized from its bare value of 1 for structureless 
QCD quarks to 0.77 for constituent quarks. 

To solve the Schr\"odinger equation for the Hamiltonian in Eq.(\ref{hamiltonian}) 
the wave functions  are expanded in a harmonic oscillator basis that includes up to 
$N=2$ excitation quanta.  The $N$ ground state and $N^*(1440)$ wave functions are given 
by a superposition of five harmonic oscillator states          
\bea
\label{wf}
|N\ \rangle& = & a_{S_S}\ |S_S\rangle+a_{S_S'}\ |S_S'\rangle
+a_{S_M}\ |S_M\rangle+a_{D_M}\ |D_M\rangle
+a_{P_A}\ |P_A\rangle\nonumber\\
|N^\star\rangle & = &a^\star_{S_S}\ |S_S\rangle+a^\star_{S_S'}\ |S_S'\rangle
+a^\star_{S_M}\ |S_M\rangle+a^\star_{D_M}\ |D_M\rangle
+a^\star_{P_A}\ |P_A\rangle ,
\eea
while for the $\Delta$ ground state we have
\be
\label{wfd}
|\Delta\rangle  =  b_{S_S}\ |S_S\rangle_{\Delta}+b_{S_S'}\ |S_S'\rangle_{\Delta}
+b_{D_S}\ |D_S\rangle_{\Delta}+b_{D_M}\ |D_M\rangle_{\Delta}.
\ee                                       
The mixing coefficients for the $N(939)$, $N^*(1440)$, and $\Delta(1232)$ wave functions 
are determined by diagonalization of the Hamiltonian in Eq.(\ref{hamiltonian})
in this restricted harmonic oscillator basis, and are given in Table~\ref{table:table1} 
(model $A$). The $N$ and $\Delta$ are mainly 
in the $S_S$ harmonic oscillator ground state, 
while the $N^*(1440)$ is mainly given by the radial excitation state denoted 
as $S'_S$. Note that the $D$ state probabilities are typically $1\%
$ or less.  
A complete  description of the wave functions in
Eqs.(\ref{wf}-\ref{wfd}) can be found in Ref.~\cite{ik2}.

\begin{table}[h!]
\begin{center}
\nobreak
\begin{tabular}[t]{c c  c  c  c  c } 
\hline
$N$ &  $a_{S_S}$ &  $a_{S'_S}$ &  $a_{S_M}$ & $a_{D_M}$ & $a_{P_A}$
\\ 
$A$ \  \ \ & 0.9585      & $-0.1475$  & $-0.2344$  & $-0.0672$ & 0.0011   \\
$B$  \  \ \ & 0.9571      & $-0.0723$  & $-0.2704$  & $-0.0753$ & 0.0005   
                                             \\     
$N^*$ &  $a^\star_{S_S}$ &  $a^\star_{S'_S}$ &  $a^\star_{S_M}$ &
 $a^\star_{D_M}$ & $a^\star_{P_A}$
\\ 
$A$ \ \ \ & 0.1689     & 0.9832  & 0.0683  & 0.0122 & $-0.0006$   \\
$B$ \ \ \ & 0.1211     & 0.9793  & 0.1604  & 0.0232 & $-0.0005$   \\
$\Delta$ &  $b_{S_S}$ &  $b_{S'_S}$ &  $b_{D_S}$ & $b_{D_M}$ & 
\\ 
$A$  \  \ \ & 0.9564      & 0.2433  & $-0.1303$  & 0.0957 & \\     
$B$  \ \ \ & 0.9283      & 0.3273  & $-0.1406$  & 0.1069 & \\      
\hline
\end{tabular}
\end{center}
\caption{ Admixture coefficients for the $N(939)$, $N^*(1440)$, and $\Delta(1232)$ 
states as defined in Eqs.(\protect{\ref{wf}-\ref{wfd}}). Model A: color-screened
confinement potential~\cite{david2}; model B: quadratic confinement potential with anharmonic
terms of Eq.(\ref{eq:qpap}).}
\label{table:table1}
\end{table}

In order to check the sensitivity of our results 
with respect to the model of confinement, we employ two confinement  
potentials. Model $A$ refers to the confinement potential in Ref.~\cite{david2}, which is 
linear at short distances and color-screened 
at large inter-quark distances as a result of quark-antiquark 
pair creation~\cite{shen}. This is our preferred choice.
In model $B$ we use a quadratic (harmonic) 
dependence on the inter-quark distance $r$ corrected by anharmonic terms
\footnote{Anharmonic terms are
needed, when using a quadratic confinement with harmonic oscillator wave functions,
in order to break the degeneracy of the harmonic oscillator states and thus 
get a reasonable excitation spectrum. }
\begin{eqnarray}
\label{eq:qpap}
V^{conf}({\bf r}_j, {\bf r}_k)
=-\b {\lambda}^c_j \cdot \b {\lambda}^c_k\ \left ( a_c\, r^2 + A+\frac{B}{r}+Cr \right). 
\end{eqnarray}
Here, the color factor $\b {\lambda}^c_j \cdot \b {\lambda}^c_k=-8/3$ for quarks in a 
color-singlet baryon, and $r=\vert {\bf r}_j - {\bf r}_k \vert$   
is the relative distance between the two quarks.
As in the case of the color screened potential~\cite{david2},
the confinement parameters together with the quark-gluon 
coupling $\alpha_s$ and the oscillator
parameter $b$ of model $B$ have been adjusted using the $N$ and $\Delta$
 mass spectrum and low energy nucleon electromagnetic properties (magnetic moments and
charge and magnetic radii). The numerical values of these parameters are given 
in Table~\ref{table:quadconf}; the corresponding parameters of model $A$ 
are listed in Ref.~\cite{david2}.
\begin{table}[h!]
\begin{tabular}{c c  c  c  c  c } 
\hline
$b$ & $\alpha_s$ & $a_c$ & $A$ & $B$ & $C$ \\
$[$ fm] & \phantom{[fm]}  & [MeV/fm$^2$] & [MeV] & [MeV\,fm] & [MeV/fm]\\
0.5844 & 1.16 & 13.141 & 14.993 & $-9.765$ & 12.152 \\
\hline
\end{tabular}
\caption{Parameters of the confinement potential in Eq.(\ref{eq:qpap}), 
harmonic oscillator parameter $b$, and gluon-quark coupling $\alpha_s$ of model $B$.} 
\label{table:quadconf}
\end{table} 

\section{Axial $N \to \Delta$ transition form factors}
\label{sec:NDelta} 

Following Llewellyn Smith~\cite{ll} one can write the most general form of the 
axial current for the $N \to \Delta$ transition describing, 
e.g., neutrino induced pion production 
on the nucleon as depicted in Fig.~\ref{fig:feynman}, as a sum of 
four axial current terms, each of which is multiplied by a Lorentz-invariant 
form factor $C_i^A(q^2)$ that depends only on the square of 
four-momentum transfer $q^2$ 
\bea
\label{axiald}
&&\bar{u}_{\Delta\nu}({\bf p'})\biggl[
\frac{C_3^A(q^2)}{M_N}\left( g^{\mu \nu} q_{\sigma}\gamma^{\sigma}-
\gamma^{\mu}q^{\nu} \right)+
\frac{C_4^A(q^2)}{M^2_N}\left( g^{\mu \nu} p'q-
p'^{\mu}q^{\nu} \right)\nonumber\\
&&\hspace{1.5cm}+C_5^A(q^2) g^{\mu \nu}
+\frac{C_6^A(q^2)}{M^2_N} q^{\mu}q^{\nu}
 \biggr] \sqrt3 \, {\bf T}^\dagger\ u({\bf p}).
\eea
Here, ${\bf T}^\dagger$ is the 1/2 to
3/2 isospin transition operator with reduced matrix element  
taken to be one. For the $n\to\Delta^+$ or
$p\to\Delta^{++}$ transitions one needs the $\bf -T^\dagger_{+1}$ component 
of the isospin transition
operator, whereas for $n\to\Delta^-$ or $p\to\Delta^{0}$ the  $\bf T^\dagger_{-1}$ 
component has to be
used~\footnote{These are the appropriate isospin components corresponding
respectively to the
quark level  axial currents $\bar \Psi_u \gamma^\mu\gamma_5 \Psi_d$
 and $\bar \Psi_d \gamma^\mu\gamma_5 \Psi_u$.}. 
In the following we use the $n\to \Delta^+$
transition for global normalization of the axial form factors. 
In Eq.(\ref{axiald}), 
$u({\bf p})$ and $u_{\Delta\nu}({\bf p}')$ are  Dirac and  Rarita-Schwinger
spinors~\cite{Eri88} 
respectively for a nucleon with three-momentum ${\bf p}$ and a $\Delta$ with 
momentum ${\bf p}'$. The four-momentum transfer $q$ is given by $q=p'-p=(q^0, {\bf q})$,
where $q^0$ is the energy transfer, and ${\bf q}$ the three-momentum transfer.
All four Adler form factors $C_i^A(q^2)$ with $i=3,\cdots 6$ 
are real from $T$ invariance. 
Before we evaluate the axial $N \to \Delta$ transition
form factors in the chiral quark model, we discuss some of their low-energy properties.

The form factor $C_5^A(q^2)$ 
is the $N \to \Delta$ analogue of the nucleon isovector axial form 
factor $g_A(q^2)$ 
\footnote{The relation between the $N \to \Delta$ axial form factor
and the Adler form factor is  $g_A^{N \to \Delta}(q^2) = \sqrt{6}\, C_5^A(q^2)$.
In the SU(6) symmetry limit the relation between the elastic $N \to N$  and
$N \to \Delta$ axial couplings is $g_A^{n \to \Delta^+}(0) = (6 \sqrt{2}/5)\, g_A^{n \to p}(0)$.}.
$PCAC$ relates  its value at $q^2=0$ to the strong $\pi N \Delta$ coupling 
constant  $g_{\pi N\Delta}(0)$ through the non-diagonal 
Goldberger-Treiman relation 
\be
\label{pcacc5}
C_5^A(0)=\frac{f_{\pi}}{\sqrt6}\frac{g_{\pi N\Delta}(0)}{M_N}.
\ee
With the empirical value for $g_{\pi N\Delta}$ at the pion mass,
as extracted from a $K$-matrix analysis of $\pi N$ scattering 
phase shifts, $g_{\pi N\Delta}(q^2=m^2_{\pi})=28.6\pm0.3$~\cite{mukho}, 
and $f_{\pi}=92.4\ $ MeV~\cite{pdg} as measured in weak pion decay,
$C_5^A(0) = 1.15\pm 0.01$ is obtained. 

The form factor $C_6^A(q^2)$ 
is the inelastic analogue of the induced pseudoscalar form factor $g_P(q^2)$ 
of the nucleon.  
In the framework of Heavy Baryon Chiral Perturbation Theory (HB$\chi$PT) 
it has been shown that at low momentum transfers $C_6^{A}(q^2)$ is given as~\cite{zhu} 
\be
\label{c6axial}
\left.C_6^{A}(q^2)\right|_{HB\chi PT}=\frac{g_{\pi N\Delta}(q^2)}{\sqrt6}f_{\pi}
\frac{M_N}{m^2_{\pi}-q^2}-\frac{1}{6\sqrt6}\,M_N\,f_{\pi}\,g_{\pi N\Delta}(q^2)\,r_A^2
+{\cal O}(q^2,m^2_{\pi}),
\ee
quite analogous to the result obtained for 
the elastic induced pseudoscalar form factor~\cite{bernard94}.  
The first term in Eq.(~\ref{c6axial}) 
is the dominant pion-pole form factor, and the second
term is the leading order non-pole term, where $r_A^2$ is the square of the axial 
$N \to \Delta$ transition radius defined as
\be
\label{axialrad}
\left.r_A^2=\frac{6}{C_5^A(0)}\frac{d\,C_5^A(q^2)}{d\,q^2}\right|_{q^2=0}.
\ee 
The weak axial $ N \to \Delta$ transition radius $r_A^2$ 
as extracted from an analysis of neutrino scattering on deuterium~\cite{kitagaki}, 
lies in the range (see below)
\be
\label{axialradnum}
r_A^2 \simeq 0.394\sim 0.477\ {\mathrm{fm}}^2.
\ee 
Combining Eq.~(\ref{c6axial}) and Eq.(\ref{axialradnum}) one gets for 
the  non--pole part of the $C_6^{A}$ form factor
\be
\left.C_6^{A,\ non-pole}(0)\right|_{HB\chi PT}\simeq -
\frac{1}{6\sqrt6}\,M_N\,f_{\pi}\,g_{\pi N\Delta}(0)\,r_A^2
= -(1.71\sim 2.07).
\ee

As discussed in Ref.~\cite{c5q3} 
the form factor $C_3^A(q^2)$ is the axial counterpart of the electric 
quadrupole ($E2$) transition form factor $G_{E2}(q^2)$~\cite{mec}, 
which is important  for determining the shape of the nucleon~\cite{hen01}.
In several analyses (see table~\ref{table:table5}) $C_3^A(q^2)=0$ is assumed.
Below, we will see that $C_3^A(q^2)$ is mainly determined
by pion-exchange currents thereby providing a unique possibility
to study the nucleon pion cloud without major interference from 
valence quark and gluon degrees of freedom. As to $C_4^A(q^2)$, in the SU(6) symmetry limit,
this form factor is connected with the scalar helicity amplitude~\cite{c5q3}, 
which in the electromagnetic case corresponds to the charge quadrupole transition 
form factor $G_{C2}(q^2)$~\cite{mec}. However, unlike the latter
$C_4^A(0) \ne 0$ in the SU(6) symmetry limit. Experimentally, both $C_3^A(q^2)$
and $C_4^A(q^2)$ are poorly known.

We now proceed and calculate all four axial $N \to \Delta$ transition
form factors. To this end we have to convert the Dirac spinors in Eq.(\ref{axiald}) 
into Pauli spinors and extract the corresponding operator structure. 
Including the normalization factors for the $N$ and $\Delta$ spinors one obtains
in the center of mass frame of the resonance
\bea
&&A^0_{N\Delta}=\sqrt3{\bf T}^\dagger\left(C_3^A(q^2)+C_4^A(q^2)\frac{M_{\Delta}}{M_N}
-C_6^A(q^2)\frac{q^0}{M_N}
\right)\sqrt{\frac{E_N+M_N}{2E_N}}\ \frac{\b{\sigma}_{N\Delta}^{[1]}\cdot {\bf q}}{M_N}
\nonumber\\
&&{\bf A}_{N\Delta}=\sqrt3{\bf T}^\dagger\Bigg\{\b{\sigma}_{N\Delta}^{[1]}\bigg\{
\left(C_3^A(q^2)\frac{M_{\Delta}-M_N}{M_N}+C_4^A(q^2)\frac{M_{\Delta}q^0}
{M_N^2}+C_5^A(q^2)
\right)\nonumber\\
&&\hspace{3.85cm}-\frac{{\bf q}^2}{3M_N^2}\left( {C_6^A(q^2)}
+{C_3^A(q^2)}\frac{2M_N}{E_N+M_N}\right)
\bigg\}\nonumber\\
&&\hspace{2.7cm}+\left[\b{\sigma}_{N\Delta}^{[1]}\otimes
{Y}^{[2]}(\hat{{\bf q}}) \right]^{[1]}\frac{{\bf q}^2}{M_N^2}
\frac{\sqrt{8\pi}}{3}\left( C_6^A(q^2)+
\frac{C_3^A(q^2)}{4}\frac{2M_N}{E_N+M_N}\right)\nonumber\\
&&\hspace{2.7cm}-\left[\b{\sigma}_{N\Delta}^{[2]}\otimes
{Y}^{[2]}(\hat{{\bf q}}) \right]^{[1]}\frac{{\bf q}^2}{M_N^2}
\frac{\sqrt{5\pi}}{\sqrt6}C_3^A(q^2)\frac{2M_N}{E_N+M_N}\Bigg\}\sqrt{\frac{E_N+M_N}{2E_N}}.
\label{blop}
\eea
The $\b{\sigma}_{N\Delta}^{[j]}$ are tensor operators of rank $j$ 
defined at the baryon level. They are normalized such that their reduced matrix elements are 
all equal to one.  Furthermore, the energy and 
three-momentum imparted by the weak probe is denoted respectively by $q^0$ and ${\bf q}$, 
$E_N=\sqrt{{\bf q}^2+M_N^2}$, and  $Y^{[2]}(\hat{\bf q})$ is a
 spherical harmonic of rank 2 with
${\hat {\bf q}} ={\bf q}/|{\bf q}|$. 
With the axial current operators of the chiral quark model~\cite{david2} 
we can calculate all Adler form factors by comparing our quark
model matrix elements with the matrix elements of the baryon level operators in 
Eq.~(\ref{blop}). Numerical results are discussed in the next section.

\subsection{$N \to \Delta$ axial form factors at $q^2=0$}

Our numerical results for the four axial $N \to \Delta$ transition 
form factors at $q^2=0$, obtained with our preferred choice for confinement (model A), are shown in Table~\ref{table:table3}.
In contrast to most model predictions 
(see Table~\ref{table:table5}), the axial coupling $C_3^A(0)$ is non-zero 
in the present approach. 
As can be seen from Table~\ref{table:table3}, the finite value for 
$C_3^A(0)$  is mainly due to pion exchange currents. Thus, this observable
may be useful for determining the relative importance of gluon and pion degrees
of freedom in the nucleon. 
There are non-zero contributions to  $C_3^A(0)$ 
already in impulse approximation. However,
 these are small compared to the $\pi$-exchange current contribution
if realistic D state probabilities  $P_D < 1\% 
$ 
consistent with the Hamiltonian (see Table~\ref{table:table1}) are employed. 

\begin{table}[h!]
\begin{center}
\nobreak
\begin{tabular}[t]{ c  c  c  c c c} 
  & Imp & Gluon &  Pion & Conf  & Total
\\ 
\hline
$C_3^A(0)$  & $-$0.0068 & $-$0.0054 & 0.049  & $-$0.0010  & 0.035\\   
$C_4^A(0)$ & $-0.56$ & 0.31 & 0.14 & $-0.15$ 
& $-0.26$\\
$C_5^A(0)$   & 0.93 & $-0.17$ & 0.14 & 0.029 & 0.93\\  
$C_6^{A,\,non-pole}(0)$ & 0.033& 0.28 &$-0.30$ &$-0.73$ &$-0.72$\\
\hline 
\end{tabular}
\end{center}
\caption{ Axial couplings $C_i^A(0)$ for the $N\to \Delta(1232)$ transition obtained with model A for
confinement (color screened confinement). 
The axial current contributions are denoted as: one-body (imp), gluon exchange current (gluon), 
pion exchange current (pion), confinement current (conf), total result (total).}
\label{table:table3}
\end{table}

For the induced pseudoscalar form factor $C_6^A(0)$, 
we evaluate only the non-pole contribution $C_6^{A, non-pole}(0)$, 
i.e. the second term in Eq.(\ref{c6axial}), which is very small in impulse approximation. 
Moreover, gluon and pion exchange current contributions 
cancel to a large extent, so that the scalar confinement current is 
the dominant contribution. This is also the case for the elastic $g_P^{non-pole}(0)$ 
axial coupling~\cite{david1}.
Our result in Table~\ref{table:table3} 
agrees in sign with the one 
predicted by Eq.(\ref{c6axial}) but it is a factor $2 \sim 3$ smaller in magnitude. 
As mentioned above, $C_6^A(0)$ is dominated by the pion-pole
contribution of Eq.({\ref{c6axial}}). The large value
$C_6^{A,\, pion-pole}(0)\approx 52$ makes the extraction of the 
small non-pole part a difficult task. 

For $C_4^A(0)$ both one-body and two-body  exchange current 
contributions are similar in size, and exchange currents
modify the result obtained in impulse approximation considerably.
Because of a cancellation of pion and confinement 
exchange currents at $q^2=0$, the gluon exchange current contribution
provides the largest correction to the impulse approximation.

The axial vector coupling $C_5^A(0)$,
which is the counterpart of the axial nucleon  coupling
 $g_A(0)$, is completely dominated by the one-body axial current (see Table~\ref{table:table3}).
As in the case of $g_A(0)$~\cite{david1}, we observe an almost complete cancellation 
between the different exchange current contributions.
Because $C_5^A(0)$ is numerically the most important axial $N \to \Delta$
coupling, we discuss it in more detail below.

\begin{table}[h!]
\begin{center}
\nobreak
\begin{tabular}[t]{ c c c c c c c} 
&\multicolumn{6}{c}{$C_5^A(0)$}\\
\hline
Quark models & 0.97 \cite{c5q1,c5q1-2} & 0.83 \cite{c5q2}
 & 1.17 \cite{c5q3} & 1.06 \cite{c5q4} 
&  0.87 \cite{c5q5} & 1.5 \cite{golli03} \\
Empirical approaches &
$1.15\pm0.23$ \cite{barish}  & $1.39\pm0.14$ \cite{c5q5} &
$1.1\pm0.2$ \cite{c5e2} & $1.22\pm0.06$ \cite{luis} & & \\  
Current algebra & 0.98 \cite{slaughter} & & & & &\\
\hline 
\end{tabular}
\end{center}
\caption{
Axial coupling $C_5^A(0)$ for the $N \to \Delta$ transition obtained by
different groups. Most of this Table has been adapted from 
Ref.~\protect\cite{luis}.}
\label{table:c5}
\end{table}

In Table~\ref{table:c5}
we give values for  $C_5^A(0)$ obtained 
by different groups using quark models~\cite{c5q5,c5q1,c5q1-2,c5q2,c5q3,c5q4,golli03} 
or empirical approaches~\cite{barish,c5q5,c5e2,luis} that fit
the $\nu p \to \Delta^{++}\mu^-$ and $\nu d \to n \Delta^{++}\mu^-$ 
cross sections. The empirical approaches employ PCAC and 
the experimental $\pi N \Delta$ coupling constant
$g_{\pi N\Delta}(q^2=m^2_{\pi})$. We also quote a 
result obtained using a broken symmetry current algebra approach to QCD~\cite{slaughter},
which does not rely on $PCAC$. 
The quark model  results for $C_5^A(0)$ are generally smaller than the value
obtained from Eq.(\ref{pcacc5}) using PCAC.
An exception is the recent calculation by Golli {\it et al.}~\cite{golli03}.  
Using a linear $\sigma$-model they get
$C_5^A(0)=1.5$, some 25\% larger than the $PCAC$ estimate in Eq.(\ref{pcacc5}).
According to the authors this comes from a meson contribution that is too large,
because in their model only mesons bind the quarks 
so that their strength is overestimated.

In our model $C_5^A(0)$ is smaller than expected from Eq.(\ref{pcacc5}) because
the axial coupling of the constituent quarks, $g_{Aq}$,
is renormalized from its bare value of $1$ to $0.77$.
While this renormalization  led to the 
correct axial couplings of the nucleon~\cite{david1,david2}, 
it is seen here to be responsible for a smaller  $C_5^A(0)$ 
than expected from $PCAC$ and the empirical $g_{\pi N\Delta}(0)$. 
Conversely, from our value for $C_5^A(0)$ in Table~\ref{table:table3} and 
Eq.(\ref{pcacc5})
we would obtain
$
g_{\pi N\Delta}(0)=22.2
$
which is only about 4/5 of the empirical strong $\pi N\Delta$ coupling constant. 
This is a large discrepancy if we think of the width of the $\Delta(1232)$ resonance. 
Our value would imply a width $\Gamma_{\Delta}\approx 70 $ MeV 
which is only 60\% of the experimental width
 $\Gamma_{\Delta}\approx 120 $ MeV.
The ratio $g_{\pi N\Delta}(0)/g_{\pi N N}(0)$ evaluated in the present model
(using PCAC to extract $g_{\pi N\Delta}(0)$ from $C_5^A(0)$)
is close to the impulse approximation result 
$\sqrt{72/25}\approx 1.7$, which is  smaller than the empirical 
ratio $g_{\pi N\Delta}(0)/g_{\pi N N}(0) = 2.1$. 

The present model is then unable to reproduce, via $PCAC$, the strong coupling constant
$g_{\pi N \Delta}$  correctly. On the other hand, it is conceivable that the ``experimental'' 
width of the $\Delta$ from which $g_{\pi N\Delta}(0)/g_{\pi N N}(0)= 2.1$ is determined contains
some non-resonant background contribution and that the true $\pi N \Delta$ 
coupling constant is actually somewhat smaller~\cite{Henley}. 
In addition, the above considerations are based
on the assumption that the non-diagonal Goldberger Treiman relation in
Eq.~(\ref{pcacc5}) is satisfied to the same accuracy as the diagonal one.
A different explanation of the failure to reproduce the $\Delta$ width is 
given in the model of Ref.~\cite{li06-1} where claims are made
that a 10\% admixture of a $qqqq\bar q$ component in the $\Delta$ wave function
could enlarge the naive three-quark model width by a factor $2\sim 3$. 
However, weak form factors have not been evaluated in this model.

\begin{table}[h!]
\begin{center}
\nobreak
\begin{tabular}[t]{ l c c  c  c  } 
  & $C_3^A(0)$  & $C_4^A(0)$ & $C_5^A(0)$ &$C_6^{A,\, non-pole}(0)$\\ 
\hline 
This work (model A)   & 0.035 & $-0.26$ & 0.93 & $-0.72$\\
Salin \cite{salin}  & 0 &$-2.7$ & 0 &0\\   
Adler \cite{adler}  & 0 &$-0.3$ & 1.2 &0  \\
Bijtebier \cite{bijtebier}& 0 &$-2.9\sim -3.6$& 1.2 & 0\\
Zucker \cite{zuck}  & 1.8 & $-1.8$ & 1.9 & 0\\
HHM \cite{c5q5}      & 0 & $-0.29\pm0.006$ &$0.87\pm 0.03$ & 0\\
SU(6) \cite{c5q3}    & 0 &$ -0.38$ & 1.17 & 0\\
Isgur-Karl \cite{c5q3} &$-0.0013$ & $-0.66$ & 1.16 & 0.032\\
Isgur-Karl 2 \cite{c5q3} &0.0008 & $-0.657$& 1.20 & 0.042\\
D-mixing \cite{c5q3} & 0.052& 0.052 & 0.813 & $-0.17$\\ 
Golli~\cite{golli03} & 0 & 0.141 & 1.53 & 1.13 \\
\hline 
\end{tabular}
\end{center}
\caption{ Results for the $N \to \Delta$ axial couplings obtained in 
different models. Most of this Table has been taken from 
Ref.\protect\cite{c5q3}.}
\label{table:table5}
\end{table}

In Table~\ref{table:table5} we compare our total results for the axial couplings 
with other model calculations.
The ingredients of the baryon 
level calculations~\cite{salin,adler,bijtebier,zuck} 
are discussed in detail in Ref.~\cite{schreiner}. 
The remaining entries in Table~\ref{table:table5} 
refer to quark model calculations. The present model is similar to the  
Isgur-Karl and Isgur-Karl 2 (IK) quark models~\cite{c5q3}. The main difference is 
that in the latter only the one-body axial current is 
taken into account (impulse approximation) and $g_{Aq}$ is kept to 1,  
whereas we include axial two-body currents and use
the renormalized axial quark coupling $g_{A q}=0.77$ 
as required by the PCAC condition.

We have already pointed out that for $C_5^A(0)$ 
axial exchange current
contributions largely cancel so that the difference 
between the IK model and the present calculation is mainly due to  
our use of the renormalized axial quark coupling of Eq.(\ref{gtrquark}). 
For $C_4^A(0)$ and $C_6^{A,\,non-pole}(0)$ the IK results are similar to our 
one-body current contribution, whereas for $C_3^A(0)$, they obtain smaller
values than our impulse approximation. 
The different wave function admixture coefficients and oscillator parameter 
used here and in Ref.~\cite{c5q3} are responsible for this discrepancy. 
In any case, the present results clearly show 
that the main contribution to $C_3^A(0)$ and
$C_6^{A,\,non-pole}(0)$ come from exchange currents.  
Both axial transition couplings are considerably larger than predicted
in impulse approximation. Also $C_4^A(0)$ is significantly affected 
by exchange currents. Our total result for $C_4^A(0)$ is in agreement 
with the one obtained by Adler (see Table~\ref{table:table5}) using dispersion relations. 

Next, we compare our results with the $D$-mixing model~\cite{c5q3}, which 
employs only one-body axial currents and $D$-wave admixture coefficients 
in the $N$ and $\Delta$ 
that have been adjusted to reproduce the nucleon axial coupling $g_A(0)$ as suggested 
by Glashow~\cite{glas}, and the electric quadrupole ($E2$) over magnetic dipole ($M1$) 
ratio in the electromagnetic $N \to \Delta$ transition. 
The net result is a large $D$-wave probability both in the nucleon 
($P_D=20\%
$) and Delta ($P_D=30\%
$) wave 
functions. 

As mentioned before, the $C_3^A$ form factor is the weak axial analogue of the
$N \to \Delta$ electric quadrupole ($E2$) transition form factor. The latter is a measure
of the deviation of $N$ and $\Delta$ shape from spherical symmetry.  
In the $D$-state mixing model the finite values obtained for $C_3^A(0)$ 
and $C_6^{A,\,non-pole}(0)$ are a reflection of  the nonspherical $N$ and $\Delta$ shape.
However, in this model the sizes and signs of the $D$-state admixtures,
and the axial current operator are not compatible with the Hamiltonian of the system, 
and, as a result, the PCAC condition is severely violated.  In a consistent theory which 
includes both one- and two-body axial currents satisfying the PCAC constraint of Eq.(\ref{pcac}), 
the nonsphericity of the $N$ and $\Delta$ 
comes mainly from the non-valence quark degrees of freedom described by 
the two-body axial currents and not from highly deformed valence quark orbits as represented 
by large $D$-state admixtures.

\subsection{$q^2$ dependence of the axial $N \to \Delta$ form factors}

In this section we discuss the  $q^2$ behavior of the axial transition form factors. 
The available experimental information comes from the analysis  of  neutrino
scattering experiments~\cite{radecky,barish,kitagaki}. Here, we refer to the
analysis done by Kitagaki 
{\it et al.}~\cite{kitagaki} where no attempt  was made to
obtain independent information on the different form factors. Instead, 
the authors used the Adler model~\cite{adler}
as developed by Schreiner and von Hippel~\cite{schreiner}. There, the 
axial form factors for the $N \to \Delta$ transition have been parametrized as
\be
C_j^A(q^2)=\frac{C_j^A(0)\ (1-a_j q^2/(b_j-q^2))}{(1-q^2/M_A^2)^2}\hspace{1cm};
\hspace{1cm} j=3,4,5
\ee
with
\bea
& & C_3^A(0)=0\hspace{.2cm},\hspace{.2cm}C_4^A(0)=-0.3\hspace{.2cm},
\hspace{.2cm} C_5^A(0)=1.2\nonumber\\
& & a_3=b_3=0 \hspace{0.2cm}, \hspace{0.2 cm} a_4=a_5=-1.21\hspace{.2cm},\hspace{.2cm} 
b_4=b_5=2\ {\mathrm {GeV}^{2}}.
\eea
In addition, it has been assumed that $C_6^A(q^2)$ is given by the pion pole contribution alone. 
The axial mass $M_A$  is the only free parameter which  has been adjusted to experiment 
with the result
\be
\label{axialmass}
M_A=1.28\ ^{+0.08}_{-0.10}\ {\mathrm {GeV}}.
\ee
In this parameterization,  the
axial radius as defined in Eq.(\ref{axialrad}) is given by
\be
r_A^2=6\left( \frac{2}{M_A^2}-\frac{a_5}{b_5} \right)
\ee
from where one gets the values in Eq.(\ref{axialradnum}).
                                                                               
\begin{figure}
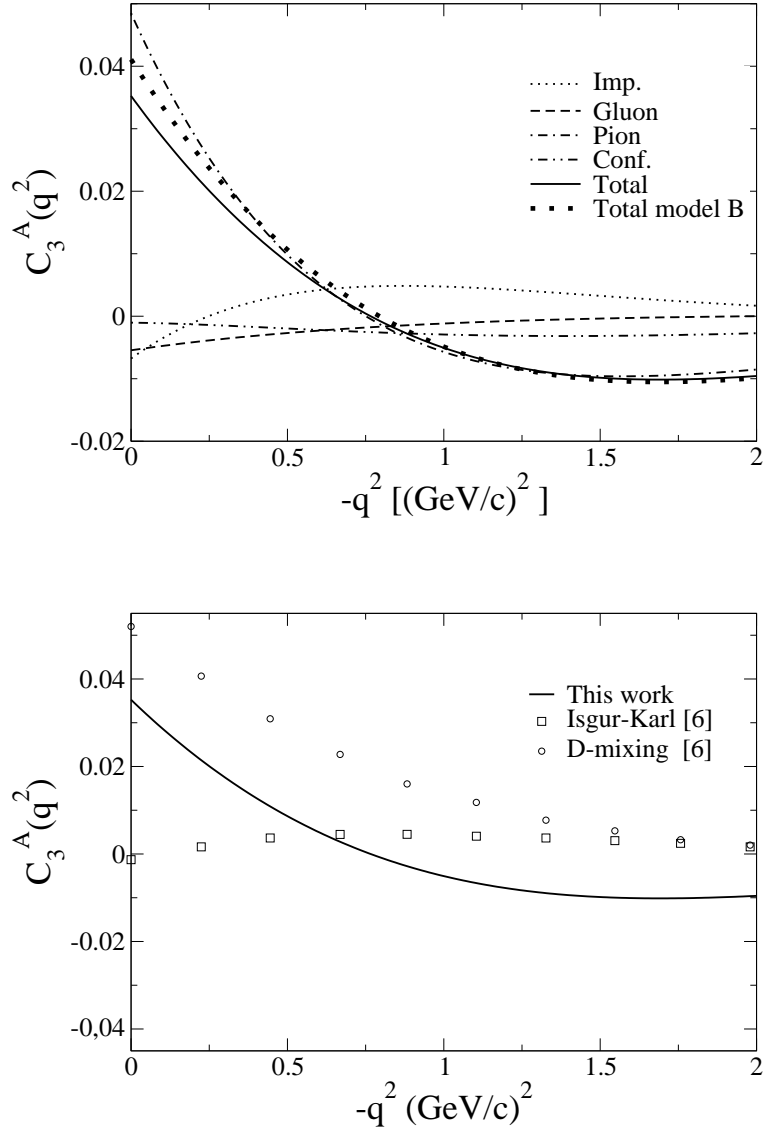

\begin{center}
\resizebox{10cm}{!}{\includegraphics{c3.eps}}\vspace{1.2cm}\\
{\resizebox{10cm}{!}{\includegraphics{c3comp.eps}}}
\caption{\label{fig:c3}
Axial form factor $C_3^A(q^2)$. Upper panel: present
model calculations with color screened confinement (model A). The individual axial current contributions are denoted as: 
impulse contribution (dotted line), gluon exchange current 
(dashed line), pion exchange current (dashed-dotted line),
confinement exchange current (dashed-double-dotted line), total result (solid line). 
The curve labelled Total model $B$ (dotted line, filled squares) represents our total result when using the confinement potential 
of Eq.(\ref{eq:qpap}). 
Lower panel: comparison with other model calculations. This work (solid line) represents our total
result obtained with our preferred choice for confinement (model A),
Isgur-Karl model~\cite{c5q3} (open squares), 
D-mixing model~\cite{c5q3} (open circles). There is no experimental information on $C_3^A(q^2)$.}
\end{center}
\end{figure}

In order to compare with experimental data and other 
theoretical calculations we evaluate the 
$q^2$ dependence of the form factors up to $q^2=2$\,(GeV/c)$^2$
with the caveat that the model may not be reliable at high momentum transfers.
For the momentum transfer dependence of the axial constituent quark coupling, 
$g_{Aq}(q^2)$, we use axial vector meson dominance 
\be
\label{avmd}
g_{Aq}(q^2)=\frac{g_{Aq}}{1-q^2/m^2_{a_1}}
\ee
with $m_{a_1}=1260$\,MeV, in analogy to the usual vector meson dominance for the
electromagnetic quark form factor~\cite{mec}. 

In Fig.~\ref{fig:c3} we show our results for $C_3^A(q^2)$.  We get  non-zero, 
though small, values which are mainly due to the pion exchange current contribution. The
form factor rapidly decreases with
$-q^2$. In the lower panel of this figure we also show the results obtained 
in the Isgur-Karl (impulse approximation) model~\cite{c5q3} leading 
to very small values over the whole $q^2$
range. In the D-mixing model calculation~\cite{c5q3}, also performed in impulse approximation,
but in this case with an unrealistic $D$-state probability in the $N$ and $\Delta$ wave
functions, larger values are obtained.
Lattice data~\cite{lattice}, not
shown in the figure,
are compatible with zero within errors, just as assumed in the experimental analysis.

\begin{figure}
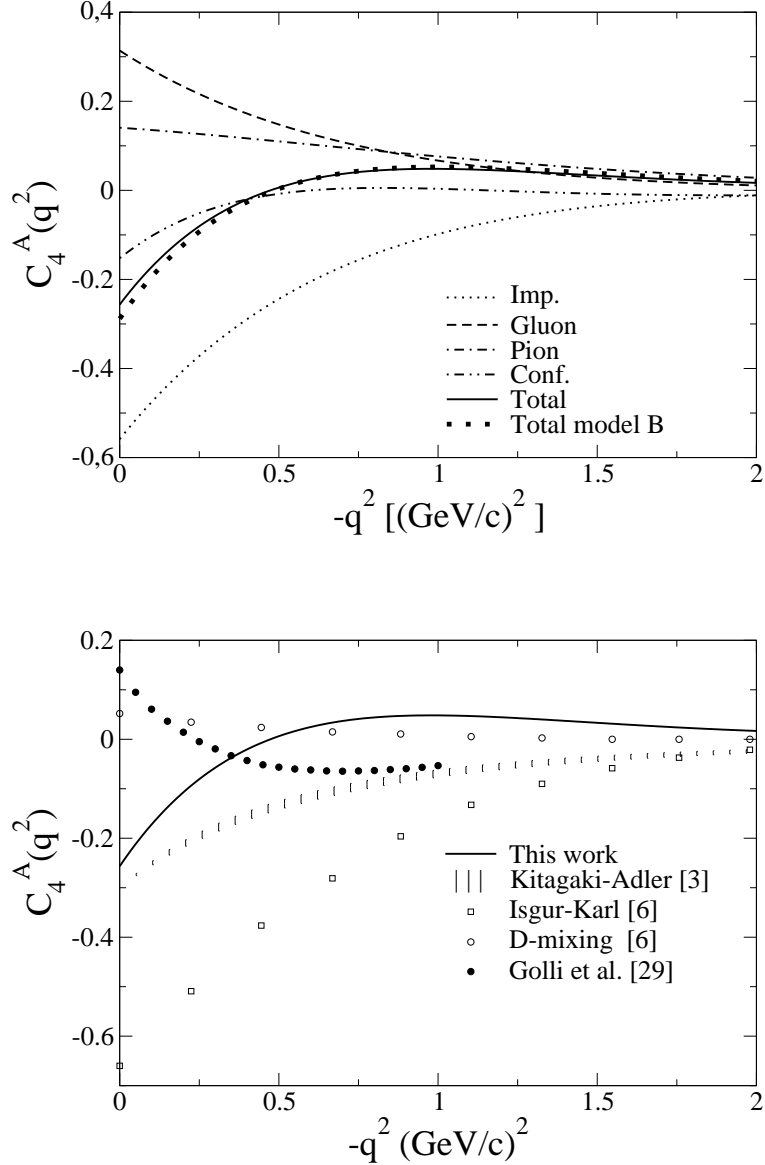

\begin{center}
\rotatebox{0}
{\resizebox{10cm}{!}{\includegraphics{c4.eps}}}\vspace{1.2cm}\\
{\resizebox{10cm}{!}{\includegraphics{c4comp.eps}}}
\caption{\label{fig:c4} 
Axial form factor $C_4^A(q^2)$. Notation as in figure~\ref{fig:c3}.
In the lower panel we also show the results of  
 the Kitagaki-Adler experimental analysis \protect\cite{kitagaki}(vertical bars),
and the ones obtained in the
linear $\sigma$-model of Golli {\it{et al.}}~\protect\cite{golli03}(black circles).}
\end{center}
\end{figure}

In Fig.~\ref{fig:c4} we plot the form factor $C_4^A(q^2)$.
Our total result starts out as expected
from the $C_4^A(q^2)=-C_5^A(q^2)/4$ relation assumed in the experimental analysis, 
but it soon deviates from it. Exchange currents are responsible for a sign change 
at around $-q^2=0.48$\,(GeV/c)$^2$. 
The results of Golli {\it et al.}~\cite{golli03},
obtained in a linear $\sigma$-model calculation, are similar in
magnitude to ours but have the opposite sign. They also show a sign 
change at approximately  $-q^2=0.24$\,(GeV/c)$^2$. 
The Isgur-Karl model calculation~\cite{c5q3} 
is very similar to our impulse contribution except for a difference in the
normalization at $q^2=0$, which, as discussed before,  
comes from the different $g_{Aq}$ values used in both calculations. 
In the D-mixing model~\cite{c5q3} very small and positive values are obtained. 

The results from the Kitagaki-Adler experimental analysis~\cite{kitagaki} are also 
shown in Fig.~\ref{fig:c4} with vertical bars. The size of the bars reflects the uncertainties in
the determination of the axial mass $M_A$ (see Eq.(\ref{axialmass})).
Quenched lattice results~\cite{lattice} display a similar behavior as
the one obtained in the calculation of Golli {\it et al.}~\cite{golli03}.
On the other hand, unquenched lattice calculations give much larger and  positive values
in the $0\le-q^2\le 2$\,GeV$^2$ region. 
Apparently,  $C_4^A(q^2)$ is very sensitive to unquenching and more statistics
is needed to draw a firm conclusion concerning its behavior~\cite{alexandrou06}.
In any case, it seems that the
assumption $C_4^A(q^2)=-C_5^A(q^2)/4$ made in the experimental analysis
is neither confirmed by quark models nor by lattice determinations.

\begin{figure}
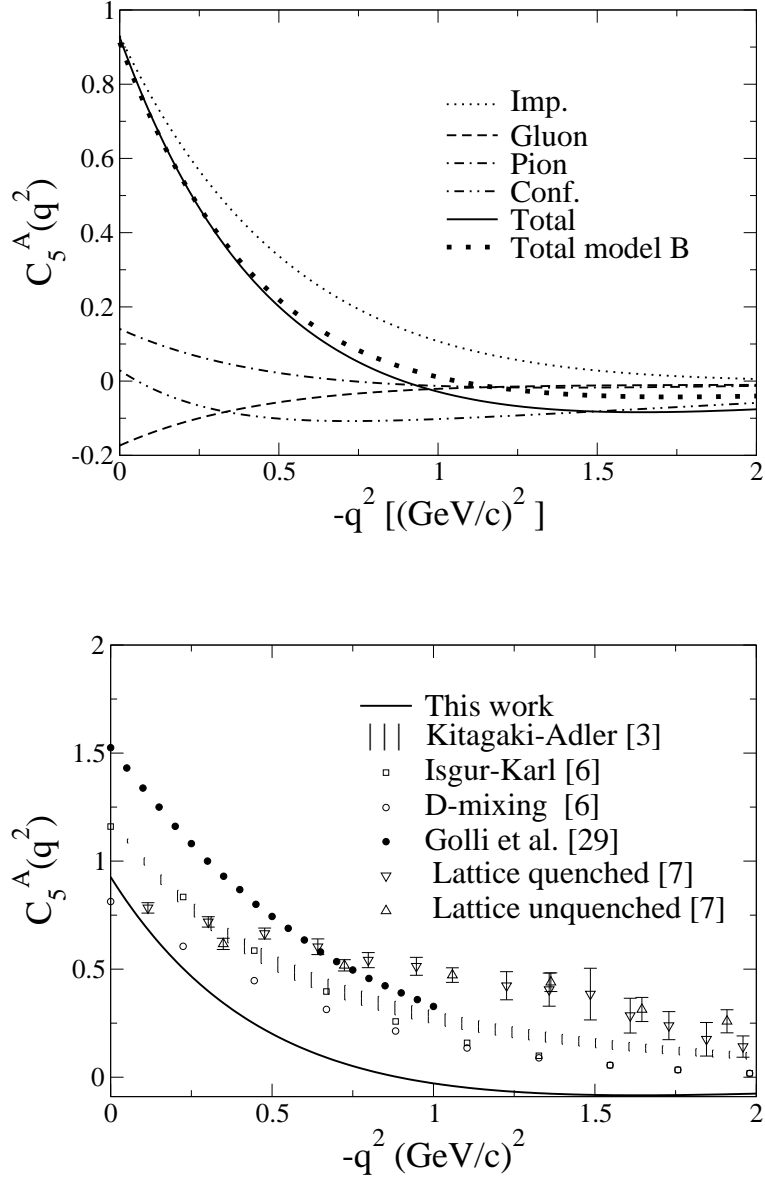

\begin{center}
\rotatebox{0}
{\resizebox{10cm}{!}{\includegraphics{c5.eps}}}\vspace{.3cm}\vspace{1cm}\\
{\resizebox{10cm}{!}{\includegraphics{c5comp.eps}}}
\caption{\label{fig:c5}
Axial form factor $C_5^A(q^2)$.
Same notation as in \hbox{Fig.~\ref{fig:c4}}. In the lower panel we also show quenched (down triangles)
and unquenched (up triangles) lattice data from Ref.~\cite{lattice}.}
\end{center}
\end{figure}
 
In Fig.~\ref{fig:c5} we present the results for $C_5^A(q^2)$. 
For this observable the impulse contribution is
dominant. In the low $q^2$  region we predict a similar behavior as the
Kitagaki-Adler analysis although with a larger slope at the origin. 
Our result for the axial radius  $\left.r_A^2\right|_{\mathrm{This\ work}}
=0.59\ $fm$^2$ 
would be closer to the one obtained in the experimental analysis if we did not include
the axial vector meson dominance factor in Eq.(\ref{avmd}). The finite axial radius
of the constituent quark~\cite{david1} contributes $r_{Aq}^2=0.147$ fm$^2$ to the axial 
transition radius.
However, in the low $q^2$ region the main difference between the present calculation and the
experimental analysis is the normalization $C_5^A(0)$.  
We obtain $C_5^A(0)=0.93$, to be compared to $C_5^A(0)=1.21$
used in the Kitagaki-Adler analysis. Again our use of the quark axial 
coupling  $g_{Aq}=0.77$ is responsible for this difference.

\begin{figure}
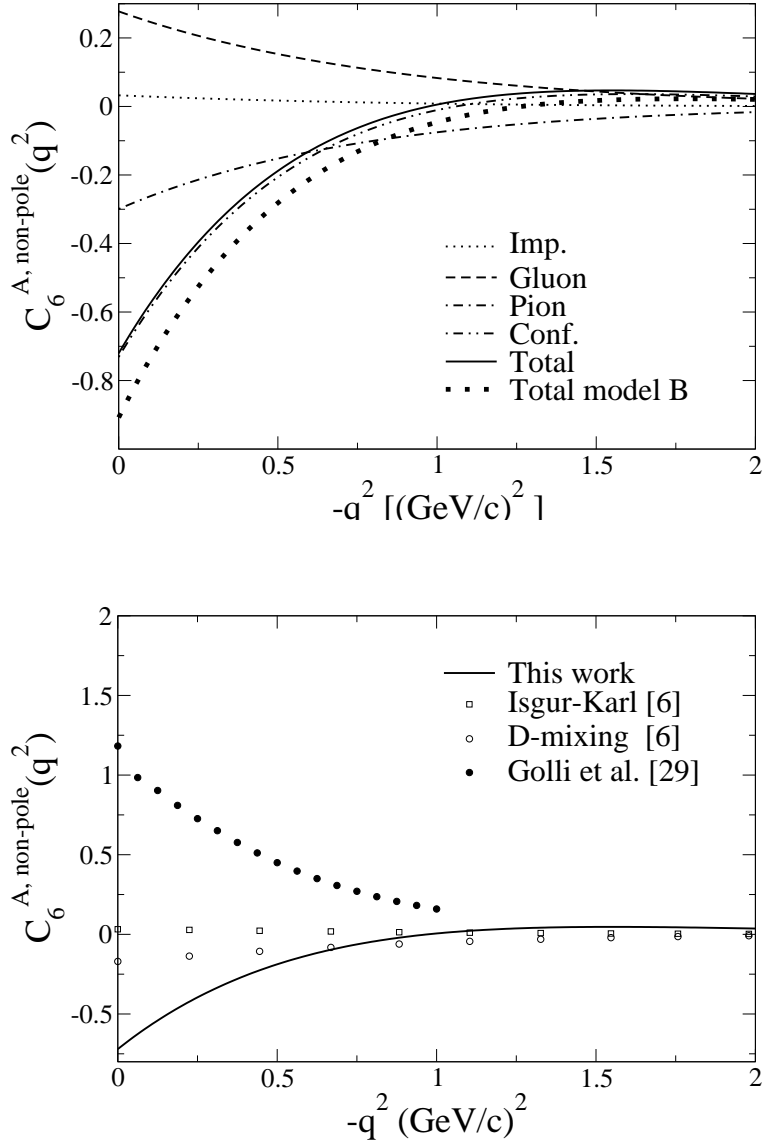

\begin{center}
\rotatebox{0}{\resizebox{10cm}{!}{\includegraphics{c6.eps}}}\vspace{1.cm}\\
\resizebox{10cm}{!}{\includegraphics{c6comp.eps}}
\caption{\label{fig:c6}
Non-pole part of the axial form factor $C_6^{A}(q^2)$. 
Notation as in Fig.~\ref{fig:c4}.}
\end{center}
\end{figure}

The $\sigma$ model calculation of Golli {\it et al.}~\cite{golli03} leads to 
a similar $q^2$ dependence at low momentum transfers as the present calculation but 
with larger $C_5^A(q^2)$ values, in particular they obtain $C_5^A(0)=1.5$. 
The Isgur-Karl model~\cite{c5q3} and our impulse contribution differ mainly
in the normalization and good agreement with the experimental analysis is obtained 
although with a smaller axial radius, 
$\left.r_A^2\right|_{\mathrm{Isgur-Karl}}\simeq 0.32\ $fm$^2$ (our estimate). 
The axial coupling $C_5^A(0)$ of the $D$-mixing model~\cite{c5q3} is similar to our
result but the $q^2$ dependence of the form factor is different leading to 
a small axial radius that we estimate to be around
$\left.r_A^2\right|_{\mathrm{D-mixing}}\simeq 0.30\ $fm$^2$. 
Quenched  lattice data from Ref.~\cite{lattice} hint at a value\footnote{Note the normalization
 for the axial form factors is different in the lattice calculation of Ref.~\cite{lattice}.
 Our $C_j^A$ are given by $C_j^A=\frac{\sqrt{2}}{\sqrt{3}}C_j^A|_{lattice}$.
  In order  to compare we multiply the lattice results by $\sqrt{2}/\sqrt{3}$ .
 }  
$\left.C_5^A(0)\right |_{Latt.\ quenched}\approx 0.86$ (our estimate from a 
linear extrapolation of their data)
with a small axial radius that we estimate as 
$\left. r_A^2\right |_{Latt.\ quenched} \approx 0.18$ fm$^2$ (our estimate).

Finally, in Fig.~\ref{fig:c6} we give our results for the form factor $C_6^{A, non-pole}(q^2)$. 
The one-body contribution is very small, while  gluon and  pion
exchange contributions cancel to a large 
extent so that the confinement exchange contribution is responsible 
for the absolute size and shape
of this form factor. The results of Golli {\it et
al.}~\cite{golli03} are similar in magnitude to the present calculation but, in contrast 
to our result they predict a positive value. This seems to contradict the findings of 
HB$\chi$PT~\cite{zhu} 
where a large negative value for $C_6^{A,\ non-pole}(0)$ has been obtained.
In the Isgur-Karl model calculation of Ref.~\cite{c5q3} small and positive values 
originating from the small $D$-wave components of the $N$ and $\Delta$ wave functions are obtained.
On the other hand,  the $D$-mixing model calculation of Ref.~\cite{c5q3}
with its large $D$-wave components leads to a negative $C_6^{A, non-pole}(q^2)$, 
although much smaller in magnitude than the present calculation. 
There is no lattice calculation of $C_6^{A, non-pole}(q^2)$. 

Comparing our results for the color screened confinement 
potential with the ones for the quadratic plus anharmonic type (Total model $B$ in 
figs.~\ref{fig:c3},~\ref{fig:c4}, and~\ref{fig:c5}) 
we conclude that there a no significant differences 
for the $C_3^A(q^2),\,C^A_4(q^2)$, and $C^A_5(q^2)$ form factors. 
In the case of $C^A_6(q^2)$ 
we observe a $25\%
$ change at small $-q^2$, with $C^A_6$ becoming more
negative and thus in better agreement with the HB$\chi$PT result~\cite{zhu}.

\section{Axial $N \to N^*(1440)$ transition form factors}
\label{sec:NRoper}

For the $n\to N^{\star+}(1440)$ transition used to normalize the form factors
\footnote{For other isospin transitions between the ground and excited state
appropriate isospin factors have to be taken into account.}, the axial current can 
be written as
\be
\label{axialr}
\bar{u}_\star({\bf p}') \left(
g^\star_A(q^2)\ \gamma^{\mu}\gamma_5
+2\frac{M_N}{m^2_{\pi}}\ g^\star_P(q^2)\ q^{\mu}\gamma_5
+g^\star_T(q^2)\ P^{\mu}\gamma_5\ \right) u({\bf p}),
\ee
where $u({\bf p})$ and $u_\star({\bf p}')$ are the Dirac spinors for the neutron
and the $N^{\star+}(1440)$ with momentum ${\bf p}$ and ${\bf p}'$ respectively, and
$P=p'+p=(P^0, {\bf P})$. 
Again the three form factors are real from $T$ invariance. 
Because the transition is not between members of the same isospin 
multiplet, invariance of strong interactions under G-parity transformations 
does not  require $g^\star_T(q^2)$ to vanish as in the case of the corresponding
elastic axial form factor~\cite{david1}.

Before we present the results of our model calculation, we first discuss 
some general properties of the $N \to N^*(1440)$ transition form factors.
As in the diagonal axial $N \to N$ transition, 
the pseudoscalar form factor $g^\star_P(q^2)$ consists 
of two terms, a pion pole and a non-pole term
\be
g^\star_P(q^2) = g^{\star \, pion-pole}_P(q^2) + g^{\star\, non-pole}_P(q^2).  
\ee
The pseudoscalar form factor $g^\star_P(q^2)$ is dominated by the pion-pole
contribution given by
\be
g^{\star,pion-pole}_P(q^2)=\frac{g_{\pi N N^\star}(q^2)}{2M_N}\,f_{\pi}\frac{m^2_{\pi}}{M_N}
\frac{M_N+M_{N^\star}}{m^2_{\pi}-q^2},
\ee
where $g_{\pi N N^\star}(q^2)$ is  
the strong $\pi N N^{\star}$ coupling
constant assuming  a pseudovector $\pi N N^\star$ coupling.
A determination of this coupling constant from the analysis of the 
Roper decay into nucleon plus pion assuming a total width of $350\ $MeV 
and a branching ratio of 65\%~\cite{pdg},
gives
\hbox{$g_{\pi NN^\star}(q^2=m^2_{\pi})=5.17$}.
In the case of the $N \to N^{\star}(1440)$ transition,
$PCAC$ relates the strong coupling constant $g_{\pi N N^{\star}}(0)$  
to the form factors  $g^\star_A(0)$ and  $g^\star_T(0)$ through 
\be
\label{PCACroper}
g^\star_A(0)+(M_{N^\star}-M_N)\cdot g^\star_T(0) =f_{\pi}\frac{g_{\pi NN^\star}(0)}{M_N}.
\ee

The operator structure at the baryon level extracted from
 Eq.(\ref{axialr}), including also the 
normalization factors for the  $N$  and $N^*(1440)$ spinors, is given in the center of
mass of the resonance by
\bea
&&A^0=\frac{\b{\sigma}\cdot {\bf q}}{E_N+M_N}
\left(-g^\star_A(q^2)-q^0\frac{2M_N}{m^2_{\pi}}\,g^\star_P(q^2)-P^0\ g^\star_T(q^2) \right)
\sqrt{\frac{E_N+M_N}{2E_N}}\nonumber \\
&&{\bf A}=\b{\sigma}\biggl[
g^\star_A(q^2) 
-\left( \frac{2M_N}{m^2_{\pi}}\ g^\star_P(q^2)-g^\star_T(q^2) 
\right)\frac{{\bf q}^2}{3(E_N+M_N)}
\biggr]\sqrt{\frac{E_N+M_N}{2E_N}}\nonumber \\
&&
\hspace{.8cm}
 + \lbrack\b{\sigma}^{[1]}\otimes {\bf q}^{[2]}\rbrack^{[1]}
\ \sqrt{\frac{5}{3}}\ \frac{1}{E_N+M_N}\left( \frac{2M_N}{m^2_{\pi}}
\ g^\star_P(q^2)-g^\star_T(q^2)  \right)\sqrt{\frac{E_N+M_N}{2E_N}}.
\eea
Here, $\b{\sigma}$ is the Pauli spin matrix operator at the baryon level and
${\bf q}^{[2]}_{m}=\lbrack{\bf q}\otimes {\bf q}\rbrack^{[2]}_{m}={\bf q}^2\sqrt{8\pi/15}\
Y^{[2]}_{m}({\hat {\bf q}})$.
These are the appropriate expressions to compare with our 
explicit constituent quark model calculation in order to extract the
axial $N \to N^{\star}(1440)$  transition form factors. 

\subsection{$ N \to N^{\star}(1440)$ axial form factors at $q^2=0$}

\begin{table}
\begin{center}
\nobreak
\begin{tabular}[t]{ c  c  c  c c c} 
  & Imp & Gluon &  Pion & Conf  & Total
\\ 
\hline
$g^\star_A(0)$  & $-0.149$ & 0.169 & $-0.169$  & $2.5\ 10^{-3}$  & $-0.148$\\   
$g^{\star,non-pole}_P(0)$ & 0.0038 & $-0.0030$ & 0.0022 & $-0.0097$ 
& $-0.0067$\\
$g^\star_T(0)$\ [$\mathrm{MeV}^{-1}$]   & $3.3\ 10^{-4}$ & $-2.8\ 10^{-5}$ &
 $1.7\ 10^{-5}$ & $1.7\ 10^{-4}$ & $4.9\ 10^{-4}$\\  
\hline 
\end{tabular}
\end{center}
\caption{ Axial couplings of the $N\to N^*(1440)$ transition obtained with model A for
confinement (color screened confinement).
The different axial exchange current contributions are denoted as in Table~\ref{table:table3}.}
\label{table:table6}
\end{table}

Our results for the different axial couplings, obtained with our preferred choice for confinement
(model
A) are given in 
Table~\ref{table:table6}. As in the case of $C_5^A(0)$ discussed above, $g^\star_A(0)$ 
is dominated
by the one-body axial current. The different two-body currents cancel to a 
large extent and the total value differs from the impulse result by 
less than 1\% . The weak axial coupling constant $g^{\star,non-pole}_P(0)$ is dominated 
by confinement exchange currents. This is similar to our result for  
the $N \to \Delta$ form factor 
$C_6^{A, non-pole}(0)$ discussed above. The axial coupling $g^\star_T(0)$ is non zero,
and receives the largest contribution from the one-body axial current. 

Using our numerical results of Table~\ref{table:table6} for $g^\star_A(0)$ and $g^\star_T(0)$, 
the evaluation of the left-hand side of Eq.(\ref{PCACroper})
yields~\footnote{Our theoretical value for the mass of the Roper is
$M^\star=1528$\ MeV.} 
\be
\label{ropercoupling}
g_{\pi NN^\star}(0)=1.43,
\ee
which is too small compared to the phenomenological value quoted above.
Although there is a  theoretical analysis~\cite{krehl} 
of the Roper width that suggests that it could be smaller, i.e., $160\ $ MeV rather 
than  $350\ $ MeV, the present $g_{\pi NN^\star}(0)$ would still be too small.
Eq.(\ref{ropercoupling}) is also at variance with Ref.~\cite{riska},
where $g_{\pi NN^\star}(0)\approx 3.5$ was obtained. On the other hand, 
their calculation is equivalent to our one-body axial current calculation, 
and thus both results for $g_{\pi NN^\star}(0)$ should agree. 
In the meantime, the correctness of our finding  in Eq.(\ref{ropercoupling}) has been
been confirmed~\cite{riskaprivate}. 
A more recent determination by the same group, 
using a Poincar\'e covariant constituent quark model with instant, point, 
and front forms of relativistic kinematics, gives values for $g_{\pi NN^\star}$ in the range
$g_{\pi NN^\star}(0)=0.71\sim 1.11$ depending on the form used~\cite{bruno} 
in agreement with our determination. 
 
It has been suggested in Ref.~\cite{jaffe} that the  $N^*(1440)$ resonance 
could be a pentaquark state that lies in the
near-ideally mixed $\overline{10}_f\oplus 8_f$ representation of SU(3)$_f$. 
This suggestion has been further supported by a QCD sum rule calculation~\cite{matheus}. 
Unfortunately, the expected width is again too small.
However,  the Roper width is not the only problem posed for the pentaquark interpretation.
There is also the problem that recent experimental results have not confirmed 
previous claims concerning the existence of pentaquark states~\cite{schumacher}. 
A different analysis argues that the Roper width can be reproduced in a model where the Roper
wave function has a 30\% admixture of a $qqqq\bar q$
component~\cite{li06-2}. As in the case of the $N \to \Delta$ transition, 
weak form factors have not yet been evaluated in this model.

\begin{figure}
\begin{center}
\rotatebox{0}{\resizebox{10cm}{!}{\includegraphics{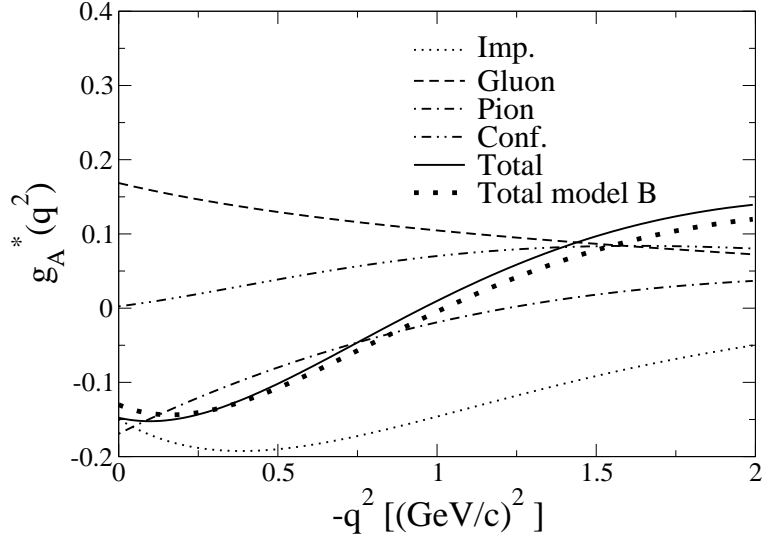}}}
\caption{\label{fig:gastar}
Axial form factor $g_A^\star(q^2)$. Notation as in the upper panel of Fig.~\ref{fig:c3}.}
\end{center}
\vspace{.5cm}
\end{figure}

\begin{figure}
\begin{center}
\rotatebox{0}{\resizebox{10cm}{!}{\includegraphics{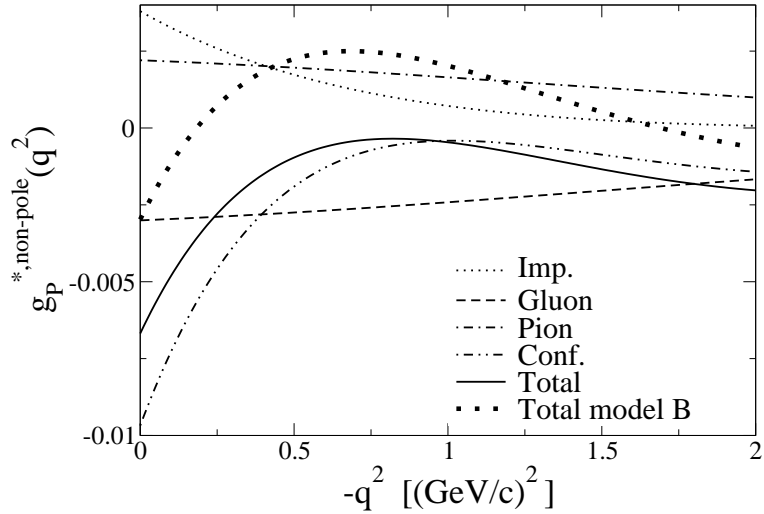}}}
\caption{\label{fig:gpstar}
Non-pole part of the  
axial form factor $g_P^\star (q^2)$. Notation as in the upper panel of Fig.~\ref{fig:c3}.}
\end{center}
\end{figure}

\subsection{$q^2$ dependence of the axial $ N \to N^{\star}(1440)$ form factors}

Next, we discuss the $q^2$ behavior of the three $ N \to N^{\star}(1440)$ form factors.
In Fig.~\ref{fig:gastar} we show  $g_A^\star(q^2)$. 
We see that gluon and pion exchange contributions cancel to a
large extent over the whole range of momentum transfers considered. 
At very low $-q^2$ the total result is 
dominated by the one-body axial current, 
while the confinement exchange current 
contribution grows as $-q^2$ increases. As a result, the minimum in the form factor 
moves to lower $-q^2$ values and we predict a sign change around $q^2 \approx 1$ GeV$^2$.
The present impulse approximation is  close in shape to the calculation of 
Ref.~\cite{bruno} using the instant and front form of relativistic
kinematics~\footnote{Note the different normalization and global sign though.}.

Our results for the $g_P^{\star, non-pole}(q^2)$ form factor are 
shown in Fig.~\ref{fig:gpstar}. Again, 
gluon and pion contributions cancel each other to a large extent. The confinement exchange
current is the dominant term at low momentum transfers but its contribution decreases in
magnitude with increasing $-q^2$.

\begin{figure}
\begin{center}
\rotatebox{0}{\resizebox{10cm}{!}{\includegraphics{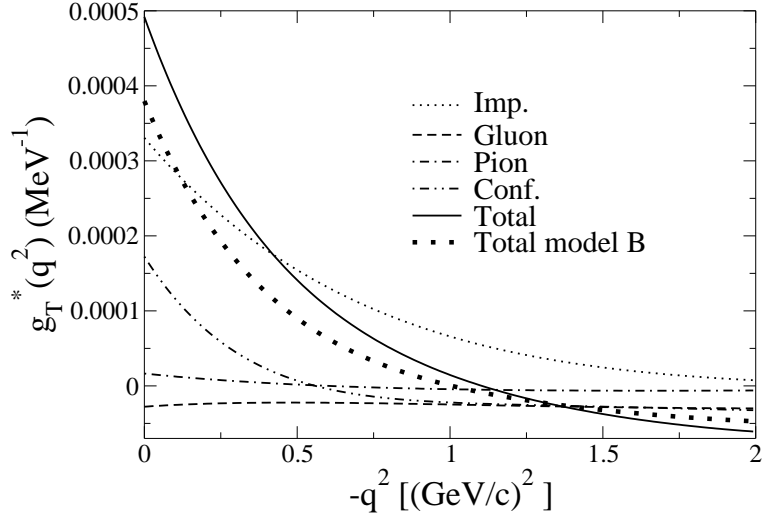}}}
\caption{\label{fig:gtstar} 
Axial form factor $g_T^\star(q^2)$. Notation as in the upper panel of Fig.~\ref{fig:c3}.}
\end{center}
\end{figure}

The form factor $g_T^\star(q^2)$ displayed in Fig.~\ref{fig:gtstar} 
is non-zero over the entire range of momentum transfers. 
Gluon and pion contributions are small and the total value at low $q^2$ is mainly given 
by the one-body axial current with confinement exchange current also playing a role.
The value for $g_T^\star(q^2)$ shows a steady decrease as $-q^2$ increases.
 
Comparison of the results with the ones obtained with the quadratic plus anharmonic
confinement potential (Total model $B$ in figs.~\ref{fig:gastar},~\ref{fig:gpstar}, 
and~\ref{fig:gtstar}) shows large changes for
$g_P^{\star,non-pole}(q^2)$ due to a decrease (in absolute value) of the confinement contribution.
For $g_A(q^2)$ and $g_T(q^2)$ the changes are not that drastic and 
the general trend of the form factors is preserved.

\section{Summary}

We have investigated the axial form factors of the weak $N \to \Delta(1232)$ 
 and $N \to N^{\star}(1440)$ 
transitions in a chiral quark model where chiral symmetry 
is introduced via a non-linear $\sigma$-model.
In contrast to previous quark model calculations, we include not only one-body
currents but also two-body axial exchange currents consistent with the 
two-body potentials in the Hamiltonian as required by the PCAC condition.

For the axial $N \to \Delta(1232)$ transition     
we find that the form factors $C_3^A(q^2)$ and $C_6^{A,\, non-pole}(q^2)$ 
are dominated by two-body currents. In particular, $C_3^A(q^2)$ is mainly determined by
pion, while $C_6^{A,\, non-pole}(q^2)$ is entirely given by the
scalar confinement exchange currents.  
Also the form factor $C_4^A(q^2)$ receives important contributions from  
axial two-body currents, mainly from the gluon exchange current. 
On the other hand, due to cancellation of the various exchange current
contributions, $C_5^A(q^2)$ is governed by the one-body axial current. 
At $q^2=0$ its magnitude is smaller than expected from $PCAC$
and the empirical strong coupling constant, $g_{\pi N\Delta}$, i.e.,
our result for $C_5^A(0)$ does not reproduce, via $PCAC$, 
the experimental value for the strong coupling constant ratio $g_{\pi N\Delta}/g_{\pi NN}$. 

For the $N \to N^{\star}(1440)$ transition, 
we find that $g_A^\star(q^2)$ is governed by the one-body axial current
but with important corrections coming from scalar confinement exchange
currents resulting in a sign change of this form factor at $q^2 \approx 1$ GeV$^2$.
At $q^2=0$ it agrees with other quark model determinations~\cite{bruno},
but it is too small to explain, via $PCAC$, 
the empirical value for the strong coupling constant $g_{\pi N N^\star}$ 
obtained from the experimental Roper resonance width.  
The form factor $g_P^{\star,\, non-pole}(q^2)$ receives the largest contribution
from two-body currents, in particular the confinement exchange current.
For $g_T^\star(0)$ we get a non-zero value mainly due to the one-body axial current but
with a 30\%  contribution coming from exchange currents.


In summary, we have found that axial two-body exchange currents play an important role 
in the weak excitation of nucleon resonances.
In particular, the axial $N \to \Delta$ transition form factor 
$C_3^A(q^2)$, which is a measure of the nonsphericity of the $N$ and $\Delta$,
is mainly determined by pion exchange currents and thus provides an 
interesting observable for studying the role of pions in the nucleon without 
interference from valence quark and gluon degrees of freedom.  
On the other hand, $C_6^{A,\, non-pole}(q^2)$ is almost exclusively determined by the 
confinement exchange current. Also the pseudoscalar 
form factor  $g_P^{\star,\, non-pole}(q^2)$ in the $N\to N^{\star}(1440)$ transition 
is largely governed by the confinement exchange current and sensitive to the confinement model. 
Further theoretical and experimental investigation of 
the axial $N \to N^{\star}$ transition form factors will undoubtedly be very useful 
for obtaining a detailed picture of nucleon structure. 

\begin{acknowledgments}
This research was supported by DGI and FEDER funds, under contracts
BFM2003-00856 and FPA2004-05616, by  Junta de Castilla y Le\'on under 
contract  SA104/04, and it is part of the EU integrated infrastructure
initiative Hadron Physics Project under contract number
RII3-CT-2004-506078.
\end{acknowledgments}

\end{document}